\documentclass[nofootinbib,superscriptaddress,twocolumn,aps,prc,10pt]{revtex4-2}

\usepackage{hyperref}
\usepackage{amsmath}
\usepackage{amssymb}
\usepackage{graphicx}
\usepackage{mathrsfs}
\usepackage{bm}
\usepackage{color}

\begin{document}
\title{Effects of center-of-mass correction and nucleon anomalous magnetic moments on nuclear charge radii}

\author{Yusuke Tanimura}
\affiliation{Department of Physics and Origin of Matter and Evolution of Galaxies (OMEG) Institute, Soongsil University, Seoul 06978, Korea}
\affiliation{Department of Physics, Tohoku University, Sendai, 980-8578, Japan}

\author{Myung-Ki Cheoun}
\affiliation{Department of Physics and Origin of Matter and Evolution of Galaxies (OMEG) Institute, Soongsil University, Seoul 06978, Korea}

\date{\today}

\begin{abstract}
Effects of the center-of-mass (CM) correction together with the nucleon electromagnetic form factors
on the nuclear charge radius are systematically studied with a relativistic Hartree-Bogoliubov model. 
Both one- and two-body parts of the CM correction are taken into account. 
It is found that the one- and two-body CM corrections, and the spin-orbit effect originating 
from the nucleon anomalous magnetic moments are all of the same order in magnitude, 
and that they give sizable impacts on the charge radius from light to heavy nuclei. 
\end{abstract}


\maketitle

\section{Introduction}

The nuclear charge radius is one of the most fundamental observables of the atomic nucleus, 
which is measured accurately by the electromagnetic probes such as electron scattering and 
atomic laser spectroscopy \cite{Ang13,Li21,Sommer22,MaEt22,Pineda21,Kuhl77,Ans86,Mar18,Mi19N,Gro20,Bar21,Good21,Kos21}.
Although the charge radius represents simply the size of the nuclear many-body system, 
it exhibits signals of the nuclear structure effects such as the shell effect \cite{Nakada19,Good21,Kos21,PeAf23,LNDS24}, 
pairing correlation \cite{Ans86,ReNa17,Mi19N,Gro20,Good21}, and deformation \cite{Kuhl77,Bar21,Cubiss23,Mun23}. 
The quantum fluctuation of the nuclear shape can also have considerable effects on the 
charge radius \cite{KERM08,Ko22}.
It is also argued that the difference of charge radii between 
a pair of mirror nuclei is correlated with the nuclear symmetry energy \cite{Br17,Mi19N,Br20,Pineda21,NaMa22,ReNa22,HLN23}. 
Therefore, the precise theoretical interpretation of the charge radius 
is intimately related to various many-body and electromagnetic effects as well as 
the understanding of nuclear force. 

Among the nuclear many-body theory, 
the mean-field model \cite{Ne82,Re89,BRM03,RS80,BR86} is suitable to study the systematic 
behaviors of the charge radius. 
It describes the nuclear many-body system in a microscopic manner 
with a universal energy density functional (EDF). 
Properties of the atomic nucleus such as binding energy, size, 
and electromagnetic moments are the basic ground-state observables that one wishes to 
describe with the model. 
An essential feature of the mean-field model is the breaking of the symmetries possessed by 
the many-body Hamiltonian. 
On the one hand, it introduces additional correlations within a single product-state 
wave function, and on the other hand, it necessitates restoration of symmetries 
or correction of the observables for the symmetry breaking \cite{RS80,BR86,Re89,BRM03}. 

The translational invariance is always violated in the mean-field model for 
finite nuclei since a many-body state is constructed as nucleons bound in a mean-field 
potential which is fixed in space. 
The center of mass (CM) of the state is localized around the potential and gives spurious 
contributions to observables. 
In principle, one should restore the symmetry by a projection method, \cite{RS80,BR86,Re89,BRM03,ScRe91}, which is numerically costly for realistic calculations.  
In most applications, the spurious effect is either neglected or 
removed in various approximate ways from the binding energy and the charge radius 
\cite{BSM83,SkM*,TM1,PK1,TyWo99,STV03,BRRM00,Co23}. 
Recently, the CM correction on the binding energy was extensively discussed in Ref. \cite{Co23} 
with a particular focus on the impact of the two-body operator part of the CM kinetic energy, 
which has been neglected in many of the existing EDFs. 
The significant effects of the two-body part on the surface-energy coefficient and the deformation energy were demonstrated \cite{BRRM00,Co23}. 

In this work, we assess the correction of the charge radius for the 
violation of translational invariance. 
The correction is made by removing the effect of the zeropoint fluctuation of the CM 
in calculating the expectation value of the squared radius. 
As in the case of the CM kinetic energy \cite{BRRM00,Co23}, 
there arise one- and two-body parts of the correction of the expectation value. 
The CM correction of the radius has often been completely neglected, 
although it is taken into account in some of the existing functionals
with the one- and two-body parts \cite{STV03,TM1} in an approximate way \cite{BSM83}, 
with only the one-body part \cite{TM1,PK1}. 
Note that, for the charge radius, the CM correction can also be taken into account in the nuclear charge form factors by an approximate projection technique \cite{ScRe91,ReNa21} 
(see also Refs. \cite{BRM03,Re89,BFST88,MiHe99,HPD09}). 
The connection between our approach and the projection method will also be discussed via a 
harmonic-oscillator model.
 
In addition to the CM correction, it is important also to consider
the electromagnetic structure of the nucleon for precise description of the charge radius, 
which is reflected in the electromagnetic form factors. 
Notice that the form factors of nucleon directly affect the 
nuclear charge-density distribution. 
In particular, the effect of the so-called ``spin-orbit'' contribution 
due to the anomalous magnetic moment of nucleon is sensitive to the shell structure, 
as has long been discussed \cite{Ber72,BFST88,Ong10,KuSu00,HoPi12,ReNa21,NOSW23}. 
Since it is an $O((v/c)^2)$ effect, it would be comparable to the CM correction of $O(1/A)$. 

Therefore, in the present work, we take into account the full CM correction of 
the charge radius, including its two-body part, together 
with the nucleon electromagnetic form factors to study systematically 
i) the contributions to the charge radius from CM correction and anomalous magnetic coupling, 
and ii) the impact of the corrections on the charge radius, 
in comparison with the experimental data. 
To be consistent with the electromagnetism formulated in a covariant way, 
it is appropriate to treat the nuclear many-body system with a relativistic theory. 
For this purpose, therefore, we employ a relativistic Hartree-Bogoliubov (RHB) model. 
It should also be noted that the significance of the relativistic nuclear mean fields 
in the anomalous magnetic coupling term has been pointed out in Refs. \cite{KuSu00,KuSu19}.

The paper is organized as follows. 
In Sec. \ref{sec:model}, we describe how the CM correction and anomalous magnetic coupling effect 
modify the calculation of charge radius. 
The analysis of the corrections and comparison with experimental data are presented 
in Sec. \ref{sec:result}. Lastly, summary and outlook is given in Sec. \ref{sec:summary}.

\section{Model}\label{sec:model}

\subsection{Relativistic Hartree-Bogoliubov model}\label{ssec:rhb}
We employ an RHB model with DDME2 parameter set \cite{ddme2} for the $ph$ channel and 
Gogny D1S interaction \cite{BGG91,YGB19} for the $pp$ channel. 
A remark on DDME2 is in order: the parameter fit to charge radii was 
made by $r_{\rm ch} = \sqrt{\langle r^2\rangle_{p} + (0.8~{\rm fm})^2}$, where $\langle r^2\rangle_{p}$ is the mean-squared (MS) radius of point-proton density distribution, 
and $(0.8\ {\rm fm})^2$ is a correction for the charge radius of the proton itself, with BCS calculations 
instead of Hartree-Bogoliubov. 
The CM correction and anomalous magnetic coupling described in the following subsections 
were not considered. 
See Refs. \cite{ddme2,ddme1,dirhb,LVPP95,SeRi02,KuRi91,AARR14,PeAfRi21} for details of the RHB model and the DDME2 parameter set. 
We impose the spherical symmetry 
and solve the RHB equations in the radial coordinate space.

\subsection{Center-of-mass correction on mean-squared radii}

The mean-square (MS) radius $\left\langle r^2\right\rangle_{p}$ of proton distribution, without CM correction, is given as 
\begin{eqnarray}
Z\left\langle r^2\right\rangle_{p} &=& \left\langle \sum_{i\in p}\bm r_i^2\right\rangle
=\int d^3r\ r^2\rho_p(\bm r), 
\label{eq:Zr2p}
\end{eqnarray}
where $\bm r_i$ is the position of the $i$th proton. 
The correction for the spurious CM contribution should be made by
\begin{eqnarray}
Z\left\langle r^2\right\rangle_{p,{\rm corr}} &=&
\left\langle \sum_{i\in p} (\bm r_i-\bm R_G)^2\right\rangle
\nonumber\\
&\equiv&
Z\left[
\left\langle r^2\right\rangle_{p}+\Delta_p^{({\rm CM}1)}+\Delta_p^{({\rm CM}2)}
\right],
\label{eq:r2corr}
\end{eqnarray}
where $\bm R_G = (1/A)\sum_{i=1}^A\bm r_i$ is the CM position of the nucleus, 
and the one- and and two-body parts of the correction, $\Delta_p^{({\rm CM}i)}$ ($i=1,2$), 
are given by
\begin{align}
\Delta_p^{({\rm CM}1)} =&
-\frac{2}{AZ} \sum_{\alpha\in p}v_\alpha^2\langle\alpha|r^2|\alpha\rangle
+\frac{1}{A^2}\sum_{\alpha}v_\alpha^2\langle\alpha|r^2|\alpha\rangle, 
\label{eq:DCM1}
\\
\Delta_p^{({\rm CM}2)} =&
+\frac{2}{AZ}\sum_{\alpha\beta\in p} (v_\alpha^2v_\beta^2-u_\alpha v_\alpha u_\beta v_\beta)|\langle\alpha|\bm r|\beta\rangle|^2
\nonumber\\
&
-\frac{1}{A^2}
\sum_{\alpha\beta} (v_\alpha^2v_\beta^2-u_\alpha v_\alpha u_\beta v_\beta)|\langle\alpha|\bm r|\beta\rangle|^2, 
\label{eq:DCM2}
\end{align}
respectively. 
$u_\alpha$ and $v_\alpha$ are the occupation amplitudes of the canonical single-particle state $\alpha$ \cite{RS80}. Notice that the summation of the first terms in 
Eqs. \eqref{eq:DCM1} and \eqref{eq:DCM2} runs over the proton states only whereas 
the one in the second terms runs over both the proton and the neutron states. 
See Appendix \ref{app:cm} for a derivation of Eqs. \eqref{eq:DCM1} and \eqref{eq:DCM2}.

\subsection{Effect of anomalous magnetic moment and finite size of nucleon}

In general, the nuclear charge form factor is given by \cite{KuSu00,KuSu19,ReNa21,PPV07,XiLi23}
\begin{align}
\tilde\rho_{\rm ch}(\bm q) 
&= 
\sum_{\tau=p,n}\int d^3r\ e^{i\bm q\cdot \bm r} 
\left[
F_{1\tau}(q^2)\rho_\tau(\bm r)
\right.
\nonumber\\
&\hspace{3.cm}
\left.
+F_{2\tau}(q^2)\rho_{\kappa\tau}(\bm r)
\right], 
\label{eq:rhoq}
\end{align}
where in the mean-field approximation
\begin{align}
\rho_\tau(\bm r) &= \sum_{\alpha\in\tau}v_\alpha^2\psi_\alpha^\dagger(\bm r)\psi_\alpha(\bm r), 
\\
\rho_{\kappa\tau}(\bm r) &= \kappa_\tau\frac{\hbar}{2mc}\bm\nabla\cdot
\sum_{\alpha\in\tau}v_\alpha^2\bar\psi_\alpha(\bm r)i\bm\alpha\psi_\alpha(\bm r), 
\label{eq:rho_k}
\end{align}
with $\psi_\alpha$ being the wave function of a canonical single-particle state $\alpha$. 
In Eq. \eqref{eq:rho_k}, $m$ is the nucleon mass, $\kappa_{p}=1.793$ and $\kappa_n=-1.913$ are 
the anomalous magnetic moments of nucleon, and $\bm\alpha=\gamma^0\bm\gamma$ is the 
usual Dirac matrix. 
The nucleon form factors $F_1(q^2)$ and $F_2(q^2)$ contain the information about the internal electromagnetic structure of nucleon. 
Note that their values at zero momentum transfer are identified as 
$F_1(0)=Q$ and $2[F_1(0)+\kappa F_2(0)]=g$, where $Q$ is the electric charge, and  
$g$ is the $g$ factor of nucleon \cite{PS95}. 
Thus they are normalized as $F_{1p}(0)=F_{2p}(0)=F_{2n}(0)=1$, and $F_{1n}(0)=0$.

The nuclear MS charge radius without the CM correction, which we denote here as $\left\langle r^2\right\rangle_{\rm ch}'$, is given by  
\begin{align}
\left\langle r^2\right\rangle_{\rm ch}'
&= -\frac{\bm\nabla^2\tilde\rho_{\rm ch}(\bm q)|_{\bm q=\bm 0}}{\tilde\rho_{\rm ch}(\bm 0)}
\nonumber\\
&=
\left\langle r^2\right\rangle_p
 + \left\langle r^2\right\rangle_{\kappa}
+ C_p + \frac{N}{Z}C_n, 
\label{eq:rch2wocm}
\end{align}
where 
\begin{eqnarray}
\left\langle r^2\right\rangle_{\kappa} &=& 
\frac{1}{Z}\sum_{\tau=p,n}\int d^3r\ r^2\rho_{\kappa\tau}(\bm r), 
\end{eqnarray}
and $C_\tau$ ($\tau=p,n$) are the constants independent of the nuclear structure,
\begin{align}
C_\tau &= -6\left.\frac{dF_{1\tau}}{dq^2}\right|_{q^2=0}
\nonumber\\
&= -6\left.\frac{dG_{E\tau}}{dq^2}\right|_{q^2=0} - \frac{3\hbar^2}{2m^2c^2}\kappa_\tau.
\label{eq:ctau}
\end{align}
Here, $G_{E\tau}=F_{1\tau}-q^2\left(\hbar/2mc\right)^2\kappa_\tau F_{2\tau}$ is 
the electric Sachs form factor \cite{ESW60,Ke02,Ke04,PPV07,GeCr11}. 
The first term in Eq. \eqref{eq:ctau} is interpreted as the MS charge radius of the nucleon itself \cite{ESW60,Hi16,Mi19}. We take the experimental values \cite{pdg22} for proton and neutron charge radii, 
\begin{align}
-6\left.\frac{dG_{Ep}}{dq^2}\right|_{q^2=0} &= (0.841\ {\rm fm})^2, 
\label{eq:rch_p}
\\
-6\left.\frac{dG_{En}}{dq^2}\right|_{q^2=0} &= -0.116\ {\rm fm}^2. 
\label{eq:rch_n}
\end{align}
Therefore, for given densities $\rho_p$ and $\rho_{\kappa\tau}$ of point nucleons, 
the momentum dependence of the form factors, or the finite-size effect, 
only adds a constant to the MS radius of point-nucleon charge distribution. 

In this work, we calculate the charge radius in the following way. 
In the RHB calculations, we take $F_{1\tau}(q^2) = F_{1\tau}(0)$ 
and $F_{2\tau}(q^2) = F_{2\tau}(0)$, i.e., we take into account the 
effect of the point-nucleon anomalous magnetic moment. The charge density is then 
given as 
\begin{eqnarray}
\rho_{\rm ch}(\bm r) 
&=& \rho_p(\bm r) 
+ \sum_{\tau=p,n}\rho_{\kappa\tau}(\bm r). 
\label{eq:jem}
\end{eqnarray}
The first term $\rho_p$ is the point-proton density distribution 
while the second term $\rho_\kappa$
describes the contributions of the anomalous magnetic couplings to the charge density. 
We refer to the latter as the ``spin-orbit'' term.
Since we use momentum-independent form factors, the finite-size effect is still neglected. 
Instead, the finite size of nucleon will be considered only at the final step to compute 
the MS charge radius by folding the resulting RHB charge density by the nucleon 
form factors, and consequently we add simply the $C_\tau$ terms to the MS radius.
We expect this is enough to the first approximation 
since the finite-size effect would give nearly the constant shift to the MS charge radius 
unless the complicated many-body effects \cite{Re89} on the nucleon form factors are 
explicitly considered. 
Note also that, according to the extra term $\sum_\tau\rho_{\kappa\tau}$ of the charge density in Eq. \eqref{eq:jem}, the equations of motion to be solved in the mean-field calculation 
for the electrostatic and the nucleon fields are modified.

With the CM correction on $\langle r^2\rangle_p$, we have for the MS charge radius, 
\begin{align}
\left\langle r^2\right\rangle_{\rm ch}
&=
\left\langle r^2\right\rangle_{p,{\rm corr}}+\left\langle r^2\right\rangle_{\kappa}
+
C_p+\frac{N}{Z}C_n
\nonumber\\
&=
\left\langle r^2\right\rangle_p
+\Delta_p^{({\rm CM}1)}+\Delta_p^{({\rm CM}2)}
 + \left\langle r^2\right\rangle_{\kappa}
\nonumber\\
&\hspace{.4cm}
+
\left(
 0.588
+ 0.011\frac{N}{Z}\ {\rm fm}^2
\right), 
\label{eq:rch2}
\end{align}
where we have substituted the numerical values for nucleon charge radii [Eqs. \eqref{eq:rch_p} and \eqref{eq:rch_n}], and the 
$3\hbar^2\kappa/2m^2c^2$ terms. 
The first term of Eq. \eqref{eq:rch2} is the MS radius of point proton, the second and third terms are 
the CM correction of the first, the fourth term is the contribution from 
the magnetic spin-orbit term, and the last term is the finite-size effect of nucleon 
introduced by the momentum-dependence of the form factors. 
Notice that the last term which is independent of the many-body wave function is 
almost constant with a weak $N/Z$ dependence.
In the present work, the CM correction of the small spin-orbit contribution 
$\langle r^2\rangle_\kappa$ is neglected. 
The root-mean-square (RMS) charge radius is defined as 
\begin{eqnarray}
r_{\rm ch} = \sqrt{\left\langle r^2\right\rangle_{\rm ch}}. 
\end{eqnarray}

\section{Results and discussions}\label{sec:result}

With the model described in the previous section, we calculate the 
charge radii of even-even nuclei in the isotope chains $^{4\mathchar`-8}$He, $^{10\mathchar`-22}$C, 
$^{12\mathchar`-28}$O, $^{36\mathchar`-56}$Ca, $^{50\mathchar`-80}$Ni, 
$^{78\mathchar`-112}$Zr, $^{100\mathchar`-148}$Sn, and $^{180\mathchar`-220}$Pb. 

For brevity, the one- and two-body CM correction, $\Delta_p^{({\rm CM}1)}$ and $\Delta_p^{({\rm CM}2)}$, and the spin-orbit term $\langle r^2\rangle_\kappa$ will be referred to as CM1, CM2, and SO, respectively. 

\subsection{Contribution of each correction}\label{ssec:dcmp}
Before making a direct comparison of calculated and measured values of the charge radius, 
we first show in Fig. \ref{fig:dr2} the contributions to the MS charge radius of the three terms, 
the SO term, $\langle r^2\rangle_\kappa$, with magenta triangles, 
the CM1 term, $\Delta_p^{({\rm CM}1)}$, with skyblue squares, 
and the CM2 term, $\Delta_p^{({\rm CM}2)}$, with purple squares. 
The sum of the three is shown by black dots. 
The gray bands in the figure show, as a reference to the size of experimental uncertainty, 
the range given by $\Delta\langle r^2\rangle({\rm exp}) \in [(r_{\rm ch}-\delta r_{\rm ch})^2-r_{\rm ch}^2:(r_{\rm ch}+\delta r_{\rm ch})^2-r_{\rm ch}^2]$, 
with $r_{\rm ch}$ and $\delta r_{\rm ch}$ being the measured value of the charge radius and the associated error, respectively. 

Remarkably, all of the three correction terms are of the same order of magnitude, and 
furthermore, each contribution as well as their sum are much larger than 
the size of experimental uncertainty except for a few cases. It implies that 
the three contributions have to be considered if one strives for precise description 
of the nuclear charge radius. 

\begin{widetext}

\subsubsection{Center-of-mass correction}\label{ssec:cm}

\begin{figure}
\includegraphics[width=\linewidth]{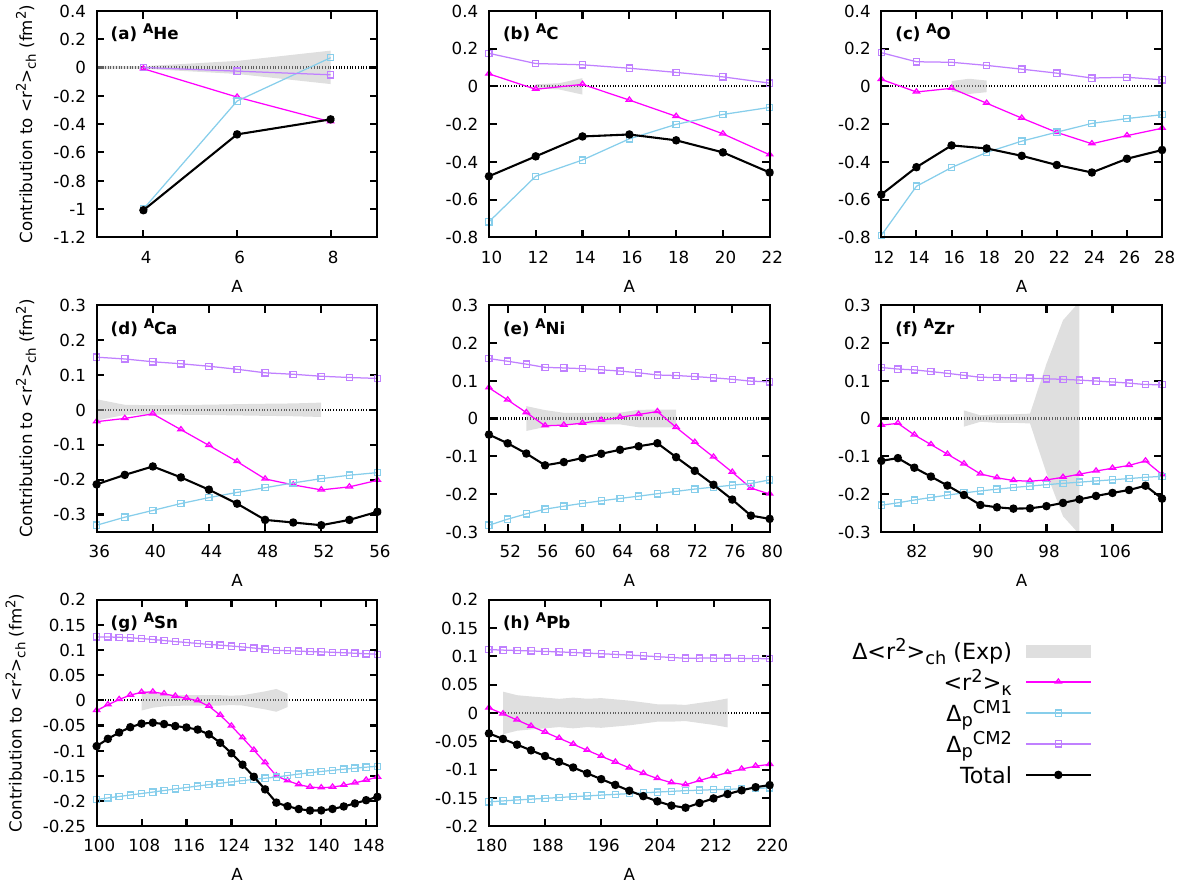}
\caption{Contributions of each correction term to the MS charge radius for (a) He, (b) C, (c) O, (d) Ca, 
(e) Ni, (f) Zr, (g) Sn, and (h) Pb isotopes. The magenta triangles show the 
anomalous magnetic contribution $\langle r^2\rangle_\kappa$, the skyblue and purple 
squares the one- and two-body CM corrections, $\Delta_p^{({\rm CM}1)}$ and $\Delta_p^{({\rm CM}2)}$, respectively, and black dots the total correction $\langle r^2\rangle_\kappa+\Delta_p^{({\rm CM}1)}+\Delta_p^{({\rm CM}2)}$. 
The gray bands show the size of experimental uncertainty 
$\Delta\langle r^2\rangle_{\rm ch}\in[(r_{\rm ch}-\delta r_{\rm ch})^2-r_{\rm ch}^2:(r_{\rm ch}+\delta r_{\rm ch})^2-r_{\rm ch}^2]$, 
with $r_{\rm ch}$ and $\delta r_{\rm ch}$ being the measured value of the charge radius and the associated error, respectively. 
The data for $^{54,56}$Ni are taken from Ref. \cite{Sommer22}, $^{58-70}$Ni from Ref. \cite{MaEt22}, and the others from Refs. \cite{Ang13,Li21}. }
\label{fig:dr2}
\end{figure}

The CM1 and CM2 terms are respectively negative and positive in most cases and rather smooth as functions of the mass number. 
Since CM1 and CM2 are $O(1/A)$ corrections, their values tend to be more substantial for 
the light nuclei but smaller and almost constant for heavy nuclei. 
Moreover, the CM2 term tend to cancel the CM1 term for heavier systems, 
representing the correct asymptotic behavior of the CM correction for $A\to\infty$, or 
infinite-matter limit. Therefore, the CM2 term should not be neglected in particular 
for heavier nuclei.

An approximation with a harmonic-oscillator model described in Appendix \ref{app:ho} is helpful 
to discuss the CM correction. As shown in Appendix \ref{app:ho}, the harmonic-oscillator model 
reproduces accurately the RHB results for Ca and heavier nuclei but only qualitatively 
for the lighter nuclei. 
With a further crude approximation in the harmonic-oscillator model, $N=Z=A/2$, 
one finds for the two-body to one-body ratio of the CM correction that 
\begin{align}
\frac{\Delta_p^{({\rm CM}2)}}{\Delta_p^{({\rm CM}1)}} = -\frac{\bar N}{\bar N+2}, 
\end{align}
where $\bar N$ is the harmonic-oscillator quantum number of the highest-occupied major shell. 
One immediately sees that the ratio tends to zero for $s$-shell nuclei 
and decreases with $A$ towards the asymptotic value $-1$ for $A\to \infty$.
One observes the similar trend in Fig. \ref{fig:dr2}. 

Now let us pick up the He isotopes showing somewhat irregular behavior, 
for which the harmonic-oscillator model may not work well because of the small mass numbers 
and the weakly-bound nucleons. 
As can be seen in Fig. {\ref{fig:dr2}}(a), $\Delta_p^{({\rm CM}1)}$ for $^8$He becomes positive, 
and $\Delta_p^{({\rm CM}2)}$ is negative for $^6$He and $^8$He. 
From Eq. \eqref{eq:DCM1}, we have for the CM1 correction, 
\begin{align}
\Delta_p^{({\rm CM}1)} &= 
\frac{1}{A}\left[
-2\left(1-\frac{Z}{2A}\right)\left\langle r^2\right\rangle_p
+\frac{N}{A}\left\langle r^2\right\rangle_n
\right]
=
\begin{cases}
\frac{1}{8}\left(-3\left\langle r^2\right\rangle_p + \left\langle r^2\right\rangle_n\right)
& {\rm for}\ ^4{\rm He}, \\
\frac{1}{18}\left(-5\left\langle r^2\right\rangle_p + 2\left\langle r^2\right\rangle_n\right)
& {\rm for}\ ^6{\rm He}, \\
\frac{1}{32}\left(-7\left\langle r^2\right\rangle_p + 3\left\langle r^2\right\rangle_n\right)
& {\rm for}\ ^8{\rm He}, \\
\end{cases}
\end{align}
where $\left\langle r^2\right\rangle_n$ is the MS radius of neutron.
Thus it is determined by the balance between negative and positive contributions from protons and neutrons, respectively.  
In the neutron-rich He isotopes, the neutron MS radius enhanced by the 
weekly-bound $p$-shell neutrons increases the CM1 term. 
See Table \ref{tb:He} for the neutron and proton MS radii and the resulting CM1 term of 
the He isotopes obtained by the RHB calculations. 
We note that the similar mechanism applies also to general near-dripline nuclei 
and that this effect is missing in the harmonic-oscillator model. 
(See also Fig. \ref{fig:dr2_cm} in Appendix \ref{app:ho} for the comparisons of the CM correction 
between the RHB and the harmonic-oscillator models. )
The negative values of the CM2 correction in $^6$He and $^8$He can be understood more simply. 
Since the two protons fill only the $s$ shell, the first term in Eq. \eqref{eq:DCM2}, 
which is positive, vanishes for the He isotopes. 
If we assume roughly that $v_{n1s_{1/2}}^2\approx 1$ and $v_{n1p_{3/2}}^2\approx (N-2)/4$ 
for the occupation probabilities of the neutron $1s_{1/2}$ and $1p_{3/2}$ states, 
respectively, 
\begin{align}
\Delta_p^{({\rm CM}2)} \approx
-\frac{2}{3}\frac{N-2}{A^2} I_{sp}^2, \ \ \ 
I_{sp} \equiv \int dr\ rG_{n1s_{1/2}}(r)G_{n1p_{3/2}}(r), 
\label{eq:DCM2He}
\end{align}
where $G_{n1s_{1/2}}(r)$ and $G_{n1p_{3/2}}(r)$ are the radial wave functions of the 
upper component of the canonical neutron $1s_{1/2}$ and $1p_{3/2}$ states, respectively.
Since $I_{sp}^2\sim 1$ fm$^2$, Eq. \eqref{eq:DCM2He} explains the small negative values 
of the CM2 term in $^6$He and $^8$He. 

\end{widetext}

We also mention here the connection of our approach to the approximate projection 
method \cite{ScRe91} via harmonic-oscillator approximation. 
Within the harmonic-oscillator model as described in Appendix \ref{app:ho}, 
the total CM correction given by Eqs. \eqref{eq:r2corr}-\eqref{eq:DCM2} satisfies
\begin{align}
\Delta_p^{({\rm CM}1)}+\Delta_p^{({\rm CM}2)} &=
-\frac{9\hbar^2}{4\langle \bm P_{\rm CM}^2\rangle}, 
\label{eq:DCMho}
\end{align}
where $\bm P_{\rm CM}$ is the CM momentum. 
On the other hand, it was shown in Ref. \cite{ScRe91} 
that the second-order Gaussian-overlap approximation to the momentum projection 
yields an effect identical to that with a harmonic-oscillator approximation. 
In their approximation, the nuclear charge form factor is corrected by 
an additional factor of $\exp\left(\frac{3\hbar^2 \bm q^2}{8\langle\bm P_{\rm CM}^2\rangle}\right)$ \cite{ScRe91,ReNa21}, 
which coincides with the CM correction of $-\frac{9\hbar^2}{4\langle \bm P_{\rm CM}^2\rangle}$ in Eq. \eqref{eq:DCMho}. 
Thus our approach yields, for heavy nuclei, approximately the same correction as the projection method, but not for light or weakly-bound nuclei for which the harmonic-oscillator model  
is not a good approximation (see Appendix \ref{app:ho}).

\begin{table}
\begin{tabular}{cccc}
\hline\hline
Nucleus & $\langle r^2\rangle_n$ & $\langle r^2\rangle_p$ & $\Delta_p^{({\rm CM}1)}$ \\ 
\hline
$^4$He & $3.91$ & $3.97$ & $-0.999$\\
$^6$He & $7.54$ & $3.87$ & $-0.237$\\
$^8$He & $9.84$ & $3.90$ & $0.0688$\\
\hline\hline
\end{tabular}
\caption{The neutron and proton MS radii, and the CM1 correction in the unit of fm$^2$ for He isotopes obtained with the RHB model. }
\label{tb:He}
\end{table}

\subsubsection{Spin-orbit effect}\label{sssec:am}

The SO effect is more sensitive than the CM corrections to the shell structure. 
As a result, the shape of the total correction for the heavier isotopes 
is determined almost by the SO effect with a shift by the CM correction. 

The behavior of $\langle r^2\rangle_\kappa$ can be qualitatively understood by  
a nonrelativistic approximation\footnote{
Note that the simple ``nonrelativistic approximation'' is only 
a poor approximation to the SO contribution in relativistic mean-field theory, as pointed out in Refs. \cite{KuSu00,KuSu19}, because of the strong relativistic potentials of hundreds of MeV, but it is still useful 
to discuss the qualitative behavior of $\langle r^2\rangle_\kappa$. 
We have found indeed that the estimates with Eq. \eqref{eq:r2k_nr}, 
$\langle r^2\rangle_\kappa~({\rm fm}^2) = -0.0422n$ for $^{4+n}$He, $= -0.0211n$ for $^{16+n}$O, 
and $=-0.0127n$ for $^{40}$Ca underestimates the RHB results by factor of $\approx 2$ in the absolute value but with the correct sign. 
}
\cite{Ong10,ReNa21,NOSW23}, 
\begin{align}
\rho_\kappa =
\frac{\kappa\hbar}{2mc}
\bm\nabla\cdot\left\langle\bar\psi i\bm\alpha\psi\right\rangle
\sim
-\frac{\kappa\hbar}{2mc}
\frac{\hbar}{mc}
\bm\nabla\cdot\bm J, 
\label{eq:rhok_nr}
\end{align}
where $\bm J$ is the nonrelativistic spin-orbit density \cite{VB72}. 
By integrating Eq. \eqref{eq:rhok_nr} with $r^2$, one finds that
\begin{align}
Z\langle r^2 \rangle_\kappa \sim 
\kappa\left(\frac{\hbar}{mc}\right)^2
\sum_{a}v_a^2(2j_a+1)\langle \bm l\cdot\bm\sigma\rangle_a
\label{eq:r2k_nr}
\end{align}
where $a$ labels a $j$ shell, and $v_a^2$ and $j_a$ are the occupation probability 
and the angular momentum of the level $a$, respectively. 
The symbol $\langle \bm l\cdot\bm\sigma\rangle_a$ is defined as 
\begin{align}
\langle \bm l\cdot\bm\sigma\rangle_a = 
\begin{cases}
+l_a   &\ {\rm for}\ j_a = l_a+1/2, \\
-l_a-1 &\ {\rm for}\ j_a = l_a-1/2,
\end{cases}
\end{align}
where $l_a$ is the orbital angular momentum of the level $a$. 
Thus neutrons in a $j_{>} = l+1/2$ ($j_{<} = l-1/2$) shell give negative (positive) 
contribution to $\langle r^2 \rangle_\kappa$, and a pair of spin-orbit doublet orbitals cancel 
each other at an $LS$-closed configuration. Since $\kappa_p$ is similar in the absolute value 
to $\kappa_n$ with the opposite sign, protons make the opposite contribution 
to $\langle r^2 \rangle_\kappa$ in $LS$-open nuclei. 
Thus $\langle r^2 \rangle_\kappa$ approximately vanishes for, e.g., 
doubly $LS$-closed or $N=Z$ nuclei. 
We illustrate here the five isotope chains for which we will show the 
isotope shifts in the next subsection. 
In the Ca isotopes shown in Fig. \ref{fig:dr2}(d), 
the increase towards zero of $\langle r^2 \rangle_\kappa$ up to $N=20$ and 
the decrease beyond is understood by the effects of neutrons filling $1d_{3/2}$ and 
$1f_{7/2}$ shells, respectively. 
In the Ni isotopes shown in Fig. \ref{fig:dr2}(e), $\langle r^2 \rangle_\kappa\approx 0$ 
at $N=Z=28$ due to the approximate isovector character of the SO effect. 
Above $N=28$, the neutrons are scattered over the $1p_{3/2}$, $1p_{1/2}$, and $1f_{5/2}$ 
states by the pairing interaction, which smoothen the variation of $\langle r^2 \rangle_\kappa$. The net increase of $\langle r^2 \rangle_\kappa$ from $N=28$ to $40$ 
is caused by the $1f_{5/2}$ neutrons. 
The large negative slope for $N>40$ is the effect of the $1g_{9/2}$ neutrons. 
In the Zr isotopes shown in Fig. \ref{fig:dr2}(f), 
$\langle r^2 \rangle_\kappa\approx 0$ at the doubly $LS$-closed $^{80}$Zr nucleus and 
decreases as the neutrons are added in the $1g_{9/2}$ shell. 
In the Sn isotopes shown in Fig. \ref{fig:dr2}(g), although the shell effect on 
$\langle r^2 \rangle_\kappa$ is smoothened by the pairing correlation, 
its decrease between $A\approx 120$ and $132$ is caused mainly by the $1h_{11/2}$ neutrons. 
Finally, in the Pb isotopes shown in Fig. \ref{fig:dr2}(h), it is again the intruder 
$1i_{13/2}$-state neutrons that mainly contribute the smooth decrease of 
$\langle r^2 \rangle_\kappa$ up to $A=208$.

Let us give a little more general discussion on the SO effect around the neutron shell closures. 
Below the larger magic numbers $N=50$, $82$, and $126$, 
the neutrons filling the intruder $j_>$ state, whose orbital angular momentum is 
larger than any levels in the shell below, mainly contribute to the decrease of 
the charge radius as approaching the magic numbers. 
Above a magic number, the decrease before is eventually compensated by filling of the 
spin-orbit partner of the intruder, but the other levels may also contribute 
at the early filling of the new shell. 
As a result, a local {\it minimum} of $\langle r^2\rangle_\kappa$ 
at or a little beyond $N=50$, $82$, or $126$ is developed.
It is not the case, however, for the lower magic numbers $N=8$ and $20$ (and $N=40$) that 
correspond to the $LS$ closures. 
In contrast to the $N\geq 50$ shell closures, 
the single-particle level below (above) an $LS$ closure is $j_<$ ($j_>$), which for the neutron case 
makes positive (negative) contribution to the charge radius, forming a local {\it maximum} 
at $N=8$, $20$, or $40$. 
Such local extrema of $\langle r^2\rangle_\kappa$ as described above are clearly observed indeed 
in Fig. \ref{fig:dr2}. 
This characteristic behavior of $\langle r^2\rangle_\kappa$ may influence the shape of the isotope 
shifts, in particular the kink structure as discussed also in Ref. \cite{NOSW23}.
See also a similar discussion based on the effect of nuclear spin-orbit force in Ref. \cite{Nakada19}.

\subsection{Comparison with experimental data}

Here we compare the following three calculations with experimental data for the charge radius. 
\begin{enumerate} 
\item The RHB calculations are done with $F_{1p}(q^2)=1$, $F_{1n}(q^2) = 0$, and $F_{2p}(q^2) = F_{2n}(q^2) = 0$, and the charge radius is calculated by $r_{\rm ch} = \sqrt{\langle r^2\rangle_p + (0.8\ {\rm fm})^2}$, denoted in Figs. \ref{fig:rch} and \ref{fig:is} as ``$+(0.8)^2$''. 

\item The RHB calculations are done with anomalous magnetic moment, i.e., 
$F_{1p}(q^2)=1$, $F_{1n}(q^2) = 0$, and $F_{2p}(q^2) = F_{2n}(q^2) = 1$, and the charge radius is calculated by Eq. \eqref{eq:rch2wocm}, denoted in Figs. \ref{fig:rch} and \ref{fig:is} as ``+FF''.

\item Same as 2. but $r_{\rm ch}$ is calculated by Eq. \eqref{eq:rch2} with the CM correction, denoted in Figs. \ref{fig:rch} and \ref{fig:is} as ``+FF+CM''.

\end{enumerate}

\begin{widetext}

\subsubsection{Absolute values of charge radii}

Fig. \ref{fig:rch} shows the calculated absolute values of the RMS charge radii $r_{\rm ch}$
in comparison with experimental data. 
The black dashed lines are the results obtained simply by $r_{\rm ch} = \sqrt{\langle r^2\rangle_p + (0.8\ {\rm fm})^2}$ without CM and SO corrections, 
and the green triangles and yellow circles are the ones obtained with only the SO and finite-size correction as in Eq. \eqref{eq:rch2wocm} 
and with the full correction as in Eq. \eqref{eq:rch2}, respectively. The experimental data 
\cite{Ang13,Li21,Sommer22,MaEt22} are shown by red squares 
with error bars. 

As was shown also in Sec. \ref{ssec:dcmp}, both CM and SO influence the 
charge radii by much more than the experimental uncertainties. 
The CM correction systematically reduces 
the charge radii. The effect is most significant for He isotopes, and less for the heavier systems. 
The SO effect is comparable to the CM correction in light nuclei and dominant 
in many of heavier nuclei. It is negative except for neutron-deficient C, O, and Ni isotopes 
and some of the Sn isotopes (see discussion in Sec. \ref{sssec:am}).

The calculated radii with the full correction of the He, C isotopes [Fig. \ref{fig:rch}(a)], and the Pb isotopes [Fig. \ref{fig:rch}(d)] tend to near the experimental values, 
while the agreements in other nuclei are deteriorated by CM and SO corrections. 
We note again that the fitting of DDME2 parameter set is done for 
$r_{\rm ch} = \sqrt{\langle r^2\rangle_p + (0.8\ {\rm fm})^2}$
without CM and SO corrections to $^{16}$O, $^{40,48}$Ca, $^{90}$Zr, $^{116,124}$Sn, 
and $^{204,208,214}$Pb nuclei \cite{ddme2}. 
It has also to be mentioned that the finite-size effect for ``+FF'' and ``+FF+CM'' values 
of the charge radius are given with different values of the nucleon sizes 
and the additional $3\kappa\hbar^2/2m^2c^2$ terms 
as compared to the one adopted in the DDME2 fit [see Eqs. \eqref{eq:rch2wocm} and \eqref{eq:rch2}].

The charge radii of the He isotopes [Fig. \ref{fig:rch}(a)] are most influenced 
by the corrections because of small $A$ and $Z$. 
Without CM and SO corrections, the charge radius is largest for $^4$He and 
is almost constant along the chain up to $^8$He. 
The slope becomes negative with the SO effect only, but the CM correction 
makes the slope positive, which follows the trend of the measured charge radii of He isotopes. 
The large staggering of $r_{\rm ch}$ in $^4$He-$^6$He-$^8$He is not reproduced. 

In the C and O isotopes, the CM correction is dominant around $N=Z$, but the 
SO effect increases as the neutrons fill the $1d_{5/2}$ state while the CM correction 
becomes smaller. As a result, the total correction is more or less constant along the chains. 
One sees a kink at $^{24}$O due to the SO effect of neutrons filling the $1d_{3/2}$ state. 

In the Ni isotopes [Fig. \ref{fig:rch}(b)], the CM correction dominates over the SO correction for $N\leq 40$. 
Above $N=40$, the strong negative SO effects of $1g_{9/2}$ neutrons suppresses the slope 
of the charge radius, forming a kink at $^{68}$Ni which was not observed in the 
recent measurement \cite{MaEt22}. 

We will discuss the Ca, Ni, Zr, Sn, and Pb isotopes in more detail with 
the isotope shifts in the next subsection.

\begin{figure}[h]
\includegraphics[width=\linewidth]{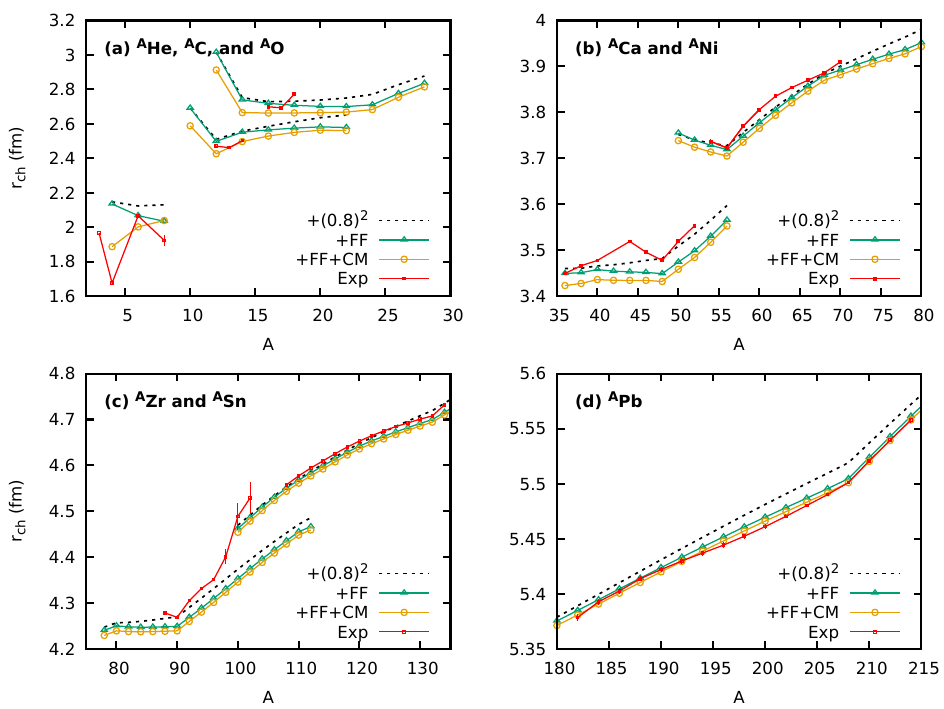}
\caption{Comparison to experimental data of the calculated charge radii 
for (a) He, C, and O isotopes, (b) Ca and Ni isotopes, (c) Zr and Sn 
isotopes, and (d) Pb isotopes.
The dashed lines show the radii calculated by $r_{\rm ch} = \sqrt{\langle r^2\rangle_p + (0.8\ {\rm fm})^2}$ without CM and SO corrections, 
the green triangles with only the SO and finite-size corrections as in Eq. \eqref{eq:rch2wocm}, 
and the yellow circles with the full correction as in Eq. \eqref{eq:rch2}. 
The experimental data \cite{Ang13,Li21,Sommer22,MaEt22} are shown by red squares with error bars. 
}
\label{fig:rch}
\end{figure}

\begin{figure}
\includegraphics[width=\linewidth]{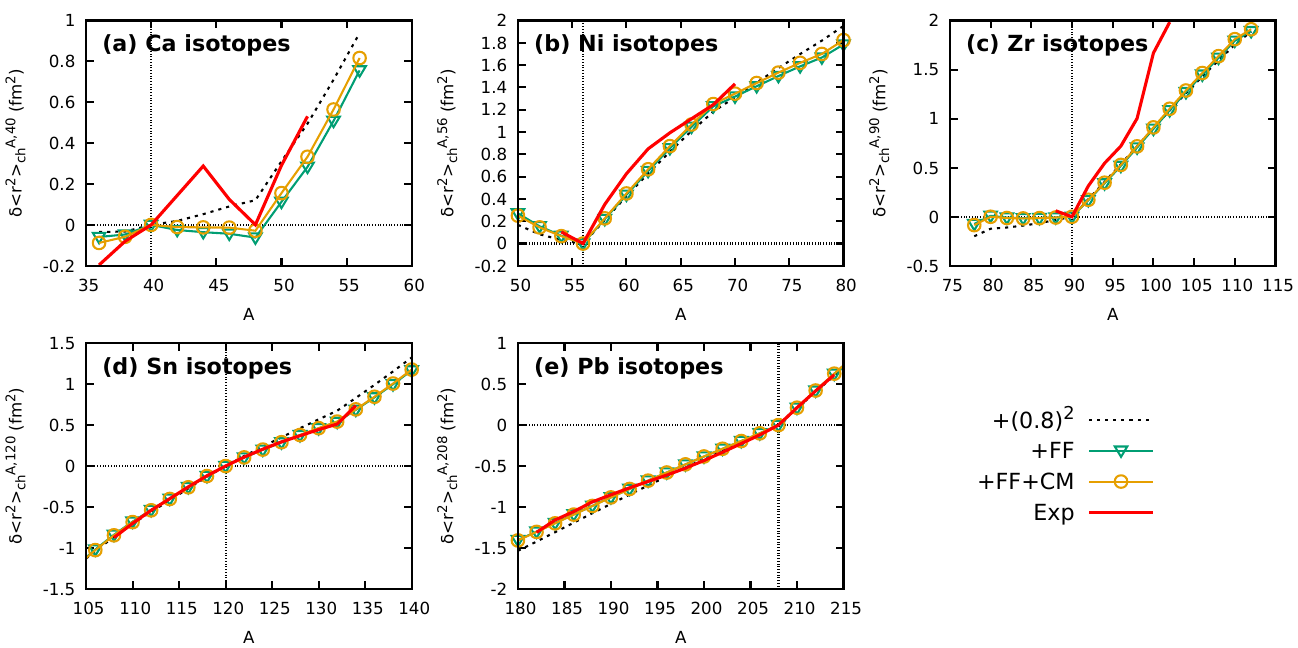}
\caption{Calculated isotope shifts compared to the experimental data for (a) Ca, 
(b) Ni, (c) Zr, (d) Sn, and (e) Pb isotopes. See caption of Fig. \ref{fig:rch} for the description of the legends. The experimental data are taken from Ref. \cite{Ang13,Li21,Sommer22,MaEt22}. }
\label{fig:is}
\end{figure}

\end{widetext}

\subsubsection{Isotopic shifts}

In order to reduce the systematic error in the calculated values of the charge radius 
coming from the above mentioned fitting procedure, we show in 
Fig. \ref{fig:is} the isotopic shifts, defined as the MS charge radius 
of an isotope $A$ relative to a reference one $A'$, 
\begin{align}
\delta\left\langle r^2\right\rangle_{\rm ch}^{A,A'}
=&
\left\langle r^2\right\rangle_{\rm ch}(A)-\left\langle r^2\right\rangle_{\rm ch}(A'). 
\end{align} 
for Ca, Ni, Zr, Sn, and Pb isotopes. 
Note that the effect of CM correction is also nearly cancelled out by the subtraction for heavier systems. 

In Ca isotopes shown in Fig. \ref{fig:is}(a), the SO effect of $1f_{7/2}$ neutrons 
drastically changes the slope of the shift between $A=40$ and $48$. 
The slight decrease of the charge radius from $^{40}$Ca to $^{48}$Ca is qualitatively 
reproduced \cite{Ber72}. 
It can also be seen that the CM correction slightly decrease the charge radius 
on $A<40$ side and increase on the other side, moderating the change of slope 
beyond $A=40$. 
The local maximum of charge radius at $^{44}$Ca and the unexpectedly large radius 
of $^{52}$Ca are not reproduced by the present calculations \cite{ReNa17,Mi19N}. 

Fig. \ref{fig:is}(b) shows the shifts in the Ni isotopes. 
A sharp kink at $^{56}$Ni observed in a recent experiment \cite{Sommer22} 
is reproduced both with and without the CM and SO corrections.
The SO effect sharpens the kink and improves the agreement with the data. 
Another kink appears at $A=68$ because of the strong SO effect of $1g_{9/2}$ neutrons. 
This kink was not observed in another recent experiment \cite{MaEt22}. 
The rapid increase of the measured charge radius above $N=28$, as in the Ca isotopes, 
forming an arch-like shape over $N=28\mathchar`-40$ is again not reproduced by the present calculations. Note that it was recently pointed out in Ref. \cite{Ko22} that 
this characteristic behavior of the charge radius between $N=28$ and $40$ is affected 
by various properties of the mean-field model such as the bulk properties, shell structure, 
and pairing correlation. 

The result for the Zr isotopes is shown in Fig. \ref{fig:is}(c).
The slope at $A<90$ region is changed mainly by the SO effect, which improves the agreement with 
the decrease of the measured charge radius from $A=88$ to $90$ [see also Fig. \ref{fig:dr2}(f)].
The large discrepancy beyond $A=90$ may be attributed to deformation effect \cite{DRHBc22}. 

In Sn isotopes shown in Fig. \ref{fig:is}(d), the decline of the slope at $A>120$ region 
is well reproduced mainly by the SO effect of $1h_{11/2}$ neutrons, 
as discussed in the previous subsection. 
The SO effect above $N=82$ shell closure is almost flat and smooth due to the scattering of 
neutrons over the shell above $N=82$ [see Fig. \ref{fig:dr2}(g)]. 
This, together with the the SO effect of $1h_{11/2}$ neutrons, leads to a kink at $A=132$ 
slightly weaker than is experimentally observed. 

Lastly, in Fig. \ref{fig:is}(e) showing the Pb isotope chain, 
the slope of the $A<208$ chain is changed by the SO effect of mainly $1i_{13/2}$ neutrons, 
which yields the constant decrease of the negative SO effect for $A<208$ [see also Fig. \ref{fig:dr2}(f)]. 
It improves the region $182\leq A\leq 192$ but slightly worsen $192\leq A\leq 206$.

We have also tried the same calculations for DDME$\delta$ parameter set \cite{ddmed}
and observed qualitatively similar effects of CM and SO corrections, 
but without a kink at $^{68}$Ni. 
It implies that the SO effect on the kink structure is sensitive 
to the proton shell structure and the proton occupation probabilities determined by the pairing correlation.
See also Ref. \cite{NOSW23}, in which a number of mean-field models are compared without 
the CM correction. Global performance studies of the DDME2 and other parameter sets 
were also done in Refs. \cite{AARR14,PeAfRi21}. 

Recently, the effect of the $\omega$-$N$ and $\rho$-$N$ tensor couplings 
in a relativistic mean-field model on the charge radii were systematically investigated \cite{LNDS24}. 
It was observed that the impact of the tensor couplings on charge radii is comparable 
to the effects considered in the present work.
The meson-nucleon tensor couplings indirectly influence the charge radius through 
its effect on the neutron spin-orbit splittings and the neutron occupation probabilities 
of the single-particle levels \cite{PeAf23}. 
The same effect was also discussed in Ref. \cite{Nakada19} with an extra 
density-dependent nuclear spin-orbit force, which leads to results resembling 
to ours for the isotope shifts in Ca, Ni, Sn, and Pb chains.  
On the other hand, the magnetic SO term in the present study, namely the photon-nucleon 
tensor coupling, is a consequence of the electromagnetic property of the nucleon, 
which directly modifies the charge density.
Note also that the SO effect is entangled with the effect of strong relativistic 
nuclear mean fields as discussed in Refs. \cite{KuSu00,KuSu19} although it is a pure electromagnetic effect, 
and that the nucleon magnetic moments or more generally the form factors 
could be modified in nuclear medium by the many-body effects and the underlying QCD quark-gluon dynamics \cite{Hen17}. 

As a final remark, the beyond-mean-field correlations other than the CM correction 
can also alter the charge radius \cite{KERM08}. 
The effects of the zeropoint quadrupole-shape fluctuation on charge radius 
was found to be as large as $\sim 0.01$ fm \cite{KERM08,Ko22}. 

\section{Summary}\label{sec:summary}

We have studied the effects of the one- and two-body CM corrections, 
and the SO term originating from the anomalous magnetic moment 
of nucleon on the nuclear charge radius. 
The former is required by the inevitable breaking of translational invariance 
in the mean-field model, 
whereas the latter is the electromagnetic property of nucleon 
affecting directly the nuclear charge-density distribution. 
The finite-size effects of nucleon from both Dirac and Pauli form factors were also included. 
We employed an RHB model with DDME2 for $ph$ channel and Gogny D1S for $pp$ channel. 

We have observed sizable impacts of each correction on the charge radius 
from light to heavy nuclei. 
The light nuclei are significantly affected by the both CM and SO corrections, 
while the heavier nuclei are much less affected by the former, as expected. 

The CM correction consists of one- and two-body parts. 
The heavier is the system, the more significant is the effect of the two-body part, 
thus it should not be neglected. We also find that the harmonic-oscillator model 
is not a good approximation in light or weakly-bound systems although 
it is nearly satisfactory for heavy systems. 

The magnetic SO effect is more sensitive to the shell structure than the CM correction. 
In particular, it leads to remarkable improvement of Sn and Pb isotope shifts for DDME2 functional. 
The SO effect also produces additional kinks at $^{24}$O and $^{68}$Ni, 
latter of which is not observed in experimental data. 

The two corrections seemingly improve also the agreement with the measured charge radii 
in very light H and C isotopes. 
Although the beyond-mean-field correlations are likely to be important in these lighter systems, 
it was shown that the present mean-field model roughly follows the trend of the measured charge radii. 

It would also be interesting to study the effects of the CM correction
on the other kinds of radius. 
More detailed analyses including those of the matter radius and neutron skin 
thickness will be reported elsewhere. 

The CM correction affects also the deformation parameters. 
The correction of the quadrupole moments can be made in a similar way as the radius 
since it is quadratic in coordinates. 
The corrections of higher moments will be much more complicated because 
there arise three-body and higher operators. 
However, it is expected that the the CM correction is 
small for the deformation parameters because cancellation of 
the correction terms would occur among different spatial directions.

\acknowledgements
We thank Toshio Suzuki for helpful discussions. 
We acknowledge support from the Basic Science Research Program of the
National Research Foundation of Korea (NRF) under Grants
No. 2021R1A6A1A03043957 and No. 2020R1A2C3006177. 

\appendix

\section{Derivation of CM correction on radius}\label{app:cm}

Derivation of Eqs. \eqref{eq:DCM1} and \eqref{eq:DCM2} is given here. 
In general, the expectation value of the square of an observable $\hat A$ is given by 
\begin{align}
\langle \hat A^2\rangle
=&
{\rm Tr}[A^2\rho] + \left({\rm Tr}[A\rho]\right)^2 
- {\rm Tr}[A\rho A\rho] - {\rm Tr}[A^*\kappa^* A\kappa], 
\end{align}
where $\rho$ and $\kappa$ are the one-body density matrix and the paring tensor, 
respectively, and $A$ is matrix representation of the operator $\hat A$. 
The first term in the right hand side is the one-body operator part of $\hat A^2$, 
while the rest is the two-body part. 
If $\hat A$ is a time-even operator, 
\begin{align}
\langle \hat A^2\rangle
=&
\sum_{\alpha} v_\alpha^2\langle\alpha|A^2|\alpha\rangle
+
\left(\sum_\alpha v_\alpha^2\langle\alpha|A|\alpha\rangle\right)^2
\nonumber\\
&
-
\sum_{\alpha\beta}(v_\alpha^2v_\beta^2
-
u_\alpha v_\alpha u_\beta v_\beta)
\left|\langle\alpha|A|\beta\rangle\right|^2, 
\label{eq:A2e}
\end{align}
where $v_\alpha$ and $u_\alpha$ are the canonical occupation amplitudes. 
Note that summations run over the time-reversal partner states pairwise. 
If $\hat A$ is time-odd, on the other hand, 
\begin{align}
\langle \hat A^2\rangle
=&
\sum_{\alpha} v_\alpha^2\langle\alpha|A^2|\alpha\rangle
\nonumber\\
&
-
\sum_{\alpha\beta}(v_\alpha^2v_\beta^2
+
u_\alpha v_\alpha u_\beta v_\beta)
\left|\langle\alpha|A|\beta\rangle\right|^2. 
\label{eq:A2o}
\end{align}
Note the opposite signs of the last terms in Eqs. \eqref{eq:A2e} and \eqref{eq:A2o}. 
Eq. \eqref{eq:A2o} applies to the expectation value of the center-of-mass kinetic energy \cite{BRRM00}. 

The proton squared radius with CM correction is given by 
\begin{align}
&
\left\langle\sum_{i\in p}(\bm r_i-\bm R_G)^2\right\rangle
\nonumber\\
=&
Z\langle r^2\rangle_p
-\frac{2}{A}\left\langle \left(\sum_{i\in p}\bm r_i\right)^2 \right\rangle
+\frac{1}{A}\left\langle \left(\sum_{i=1}^A\bm r_i\right)^2 \right\rangle, 
\end{align}
where $\bm R_G=(1/A)\sum_{i=1}^A\bm r_i$.
The second and third terms, which are the CM correction terms, 
can be computed by \eqref{eq:A2e} to obtain Eqs. \eqref{eq:DCM1} and \eqref{eq:DCM2}.

\section{Harmonic-oscillator model}
\label{app:ho}

In this appendix, we give an analytic estimate, similar to the one in Ref. \cite{BSM83}, 
of the charge radius and the 
CM correction terms with a harmonic-oscillator (HO) model, 
and compare them with the experimental data and the RHB results. 
A connection of our approach with an approximate projection method \cite{ReNa21,ScRe91}
is also demonstrated at the end. 

Let us consider particles with $\nu$ intrinsic degrees of freedom (spin and/or isospin)
filling HO shells up to the one of $\bar N$ quanta. The total number of particles $N_p$ is given by 
\begin{align}
N_p &= \sum_{n=0}^{\bar N} \nu\frac{1}{2}(n+1)(n+2)
=\frac{\nu}{6}(\bar N+1)(\bar N+2)(\bar N+3). 
\label{eq:Np}
\end{align}
The squared radius within the HO model is given by 
\begin{align}
\sum_{\alpha}v_\alpha^2\langle\alpha|r^2|\alpha\rangle
=&
\frac{\hbar}{m\omega}\frac{\nu}{8}
(\bar N+1)(\bar N+2)^2(\bar N+3)
\nonumber\\
=&
\frac{3}{4}\frac{\hbar}{m\omega}N_p(\bar N+2), 
\label{eq:ho1}
\end{align}
where $\hbar/m\omega$ is the squared oscillator length which will be determined later. 
For the CM2 term, we need to compute 
$\sum_{\alpha\beta} v_\alpha^2v_\beta^2|\langle\alpha|\bm r|\beta\rangle|^2. $
Notice that we neglect the $uvuv$ term coming from the pairing tensor 
since it is only effective near the Fermi surface and much smaller than 
the $v^2v^2$ term being a bulk effect. 
Using the HO matrix element of $\bm r$, one obtains
\begin{align}
\sum_{\alpha\beta}v_\alpha^2v_\beta^2|\langle\alpha|\bm r|\beta\rangle|^2
=&
\frac{\nu}{8}\frac{\hbar}{m\omega}\bar N(\bar N+1)(\bar N+2)(\bar N+3)
\nonumber\\
=&
\frac{3}{4}\frac{\hbar}{m\omega}N_p\bar N.
\label{eq:ho2}
\end{align}

The real solution for the algebraic equation \eqref{eq:Np} is 
\begin{align}
\bar N+2 = f_\nu(N_p)^{1/3} + \frac{1}{3f_\nu(N_p)^{1/3}}, 
\label{eq:Nbar}
\end{align}
where
\begin{align}
f_\nu(N_p) =&
\sqrt{\left(\frac{3N_p}{\nu}\right)^2-\frac{1}{27}} + \frac{3N_p}{\nu}. 
\end{align}
It follows from Eqs. \eqref{eq:ho1}, \eqref{eq:ho2}, and \eqref{eq:Nbar} that 
\begin{align}
\sum_{\alpha}v_\alpha^2\langle\alpha|r^2|\alpha\rangle
=
\frac{3}{4}\frac{\hbar}{m\omega}
N_p\left[f_\nu(N_p)^{1/3} + \frac{1}{3}f_\nu(N_p)^{-1/3}\right], 
\end{align}
and
\begin{align}
&
\sum_{\alpha\beta}v_\alpha^2v_\beta^2|\langle\alpha|\bm r|\beta\rangle|^2
\nonumber\\
=&
\frac{3}{4}\frac{\hbar}{m\omega}
N_p\left[f_\nu(N_p)^{1/3} + \frac{1}{3}f_\nu(N_p)^{-1/3}-2\right]. 
\end{align}
Notice that these two expressions have the same limiting value for $\bar N\to\infty$. 

The neutron, proton, and matter MS radii are then given by 
\begin{align}
\langle r^2\rangle_n
=&
\frac{3}{4}\frac{\hbar}{m\omega_n}\left[f_2(N)^{1/3} + \frac{1}{3}f_2(N)^{-1/3}\right], 
\\
\langle r^2\rangle_p
=&
\frac{3}{4}\frac{\hbar}{m\omega_p}\left[f_2(Z)^{1/3} + \frac{1}{3}f_2(Z)^{-1/3}\right], 
\\
\langle r^2\rangle_m
=&
\frac{1}{A}\left(N\langle r^2\rangle_n+Z\langle r^2\rangle_p\right), 
\end{align}
respectively. 
Here we allow the oscillator parameter different between neutron and proton. 
The CM1 term is given by substituting the above expressions into Eq. \eqref{eq:DCM1}, 
and the CM2 term is given as
\begin{align}
\Delta^{({\rm CM}2)}_p
=&
-\frac{3}{4}
\frac{\hbar}{m\omega_p}
\frac{Z}{A^2}
\left(1-\frac{2A}{Z}\right)
\nonumber\\
&\times
\left[f_2(Z)^{1/3} + \frac{1}{3}f_2(Z)^{-1/3}-2\right]
\nonumber\\
&
-
\frac{3}{4}
\frac{\hbar}{m\omega_n}
\frac{N}{A^2}
\nonumber\\
&\times
\left[f_2(N)^{1/3} + \frac{1}{3}f_2(N)^{-1/3}-2\right]. 
\end{align}
Note that we treat neutrons and protons separately and 
do not set $N=Z=A/2$ as is done normally in estimations of this kind \cite{RS80,BR86,BSM83}. 

We have made no approximation so far within the HO model. 
Now we make the only ansatz for the oscillator parameter $\hbar/m\omega$ 
that remains yet to be determined, 
\begin{align}
\frac{3}{4}\frac{\hbar}{m\omega_n}
=
\frac{3}{4}\frac{\hbar}{m\omega_p}
=
\left(\frac{2}{3}\right)^{1/3}
\frac{3}{5}r_0^2A^{1/3}, 
\end{align}
with $r_0\approx 1.2$ fm. This corresponds to 
approximating the oscillator frequency by $\hbar\omega\approx 41 A^{-1/3}$ MeV \cite{RS80,BR86}. 
One could also consider $(N,Z)$-dependent oscillators different 
between neutron and proton, but we take the simplest assumption with a single parameter $r_0$. 
Under this ansatz, the total CM correction simplifies to 
\begin{align}
\Delta_p^{({\rm CM}1)}+\Delta_p^{({\rm CM}2)}
=&
-\frac{3}{4}\frac{\hbar}{m\omega}\frac{2}{A}
\label{eq:DCM}
\\
=&
-\left(\frac{2}{3}\right)^{1/3}
\frac{6}{5}r_0^2A^{-2/3}, 
\end{align}
which coincides with the expression for the CM correction adopted in TM1 parametrization \cite{TM1}. 

\begin{figure}
\includegraphics[width=\linewidth]{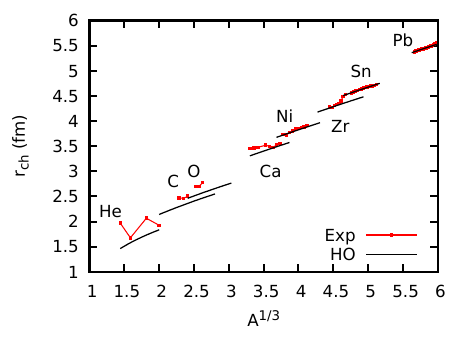}
\caption{Charge radii of H, C, O, Ca, Ni, Zr, Sn, and Pb isotopes estimated with the 
harmonic-oscillator model with $r_0=1.23$ fm compared to experimental data. 
The HO model results and the experimental data are shown by black solid curves 
and red squares, respectively. }
\label{fig:rch-HO}
\end{figure}

In Fig. \ref{fig:rch-HO} is shown the the HO-model estimate of 
the charge radius in comparison with experimental data. 
The estimate is made by substituting the HO-model values of $\langle r^2\rangle_p$ and $\Delta_p^{({\rm CM}i)}$ $(i=1,2)$
into Eq. \eqref{eq:rch2} but without the $\langle r^2\rangle_\kappa$ and the constant terms. 
We take $r_0=1.23$ fm fitted to the measured charge radii 
of Pb and Sn isotopes. 
One can see that the HO model with a single parameter $r_0$ reproduces 
the measured charge radii reasonably well from light to heavy nuclei. 
In particular, the present HO model closely follows the deviation of 
the measured values from the simple empirical formula $R = r_0A^{1/3}$. 
Although the model does not take into account the Coulomb effect, shell effect, 
deformation, etc., it captures the rough $(N,Z)$ dependence of the radius. 

\begin{figure}
\includegraphics[width=\linewidth]{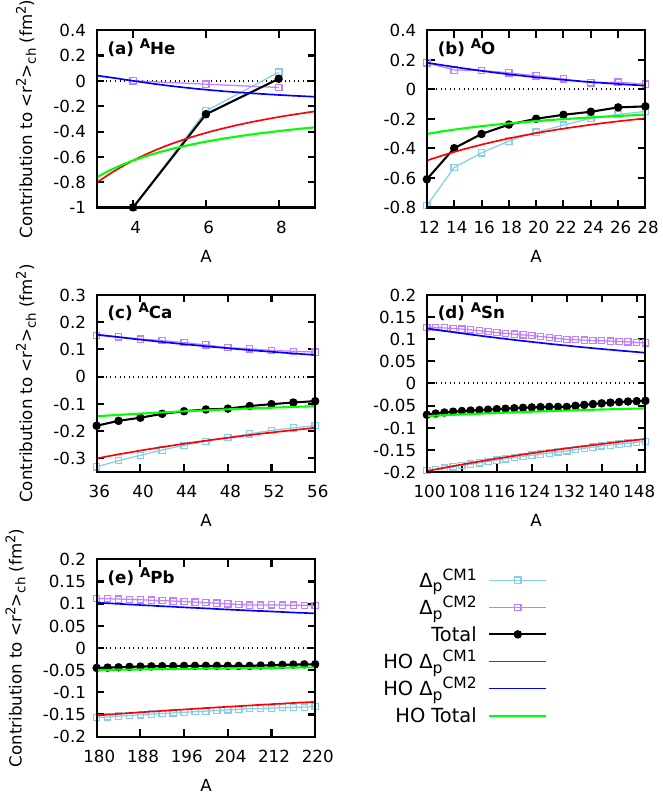}
\caption{Comparisons of the CM correction between the HO model ($r_0=1.23$ fm) and RHB results 
for (a) He, (b) O, (c) Ca, (d) Sn, and (e) Pb isotopes. 
The skyblue and purple squares show the RHB results for one- and two-body CM corrections, respectively, and black dots the total correction $\Delta_p^{({\rm CM}1)}+\Delta_p^{({\rm CM}2)}$. 
The HO model estimates for one- and two-body CM correction and their sum are shown by 
red, blue, and green curves, respectively. }
\label{fig:dr2_cm}
\end{figure}

Using the same value of $r_0$ adjusted to the measured charge radii, 
we also compare the HO model with RHB results.  
In Fig. \ref{fig:dr2_cm}, we show the comparison of $\Delta_p^{({\rm CM}1)}$ and $\Delta_p^{({\rm CM}2)}$ between the RHF calculations and the HO estimates. 
It is found that the HO model gives only qualitative estimates for H and O isotopes, 
while the agreement is nearly satisfactory for Ca, Sn, Pb isotopes.
There are two reasons of the discrepancies in the light isotopes. 
First, the enhancement of the radius by the weekly-bound 
nucleons in near-dripline nuclei is not taken into account in the HO model, 
as discussed in Sec. \ref{ssec:cm}. 
Second, the simple assumption of $\hbar\omega\approx 41A^{-1/3}$ MeV may not be
good for the very light nuclei. 

The CM correction for the kinetic energy can also be computed in 
the HO model as 
\begin{align}
\langle \bm P_{\rm CM}^2\rangle
&\approx
\sum_\alpha v_\alpha^2 \langle\alpha|p^2|\alpha\rangle
-
\sum_{\alpha\beta} v_\alpha^2v_\beta^2 |\langle\alpha|\bm p|\beta\rangle|^2
\nonumber\\
&=
\frac{3}{4}\hbar^2\frac{m\omega}{\hbar}\cdot 2A. 
\label{eq:PCM2ho}
\end{align}
From Eqs. \eqref{eq:DCM} and \eqref{eq:PCM2ho}, one finds the approximate relationship 
of the CM correction between MS charge radius and kinetic energy, 
\begin{align}
\Delta_p^{({\rm CM}1)}+\Delta_p^{({\rm CM}2)} &=
-\frac{9\hbar^2}{4\langle \bm P_{\rm CM}^2\rangle}. 
\end{align}
This expression is consistent with the CM correction adopted in Ref. \cite{ReNa21}
with an approximate projection method \cite{ScRe91} giving an additional 
factor of 
$\exp\left(\frac{3\hbar^2 \bm q^2}{8\langle\bm P_{\rm CM}^2\rangle}\right)$ to the nuclear charge form factor.

\end{document}